\documentclass[11pt]{article}      
\author{Kris Krogh \\
Neuroscience Research Institute\\ University of California, Santa
Barbara, CA 93106, USA\\ email: k\_krogh@lifesci.ucsb.edu}

\title
{Origin of the Blueshift in Signals from \linebreak Pioneer 10 and
11}

\date{June 21, 2005}

\begin{document}

\maketitle

\begin{abstract}

\normalsize A previous paper~\cite{kk} introduced a
quantum-mechanical theory of gravity, and showed it agrees with
the standard experimental tests of general relativity. Doppler
tracking signals returned by the Pioneer 10 and 11 space probes
offer an additional test. Analysis by Anderson, Laing, Lau, Liu,
Neito and Turyshev~\cite{alllnt} finds a persistent blueshift,
equivalent to an extra acceleration of the probes toward the Sun.
While unexplained by general relativity or prevailing cosmology,
it's shown this effect is predicted by the quantum-mechanical
alternative.

\end{abstract}

\vfill\eject

\section{Introduction}

NASA's Pioneer 10 probe was launched in 1972, and sent the first
close-up pictures of Jupiter.  Pioneer 11, launched the following
year, showed us Saturn. In addition to those revealing images, they
sent a puzzle: Doppler tracking signals returned by both probes
indicated an anomalous acceleration toward the Sun~\cite{alllnt}.
Pioneer 10 crossed Pluto's orbit in 1983 and continued sending data
until two years ago, when it was 82 AU from the Sun. Pioneer 11
relayed tracking signals until an electronics unit failed in 1990,
at a distance of 30 AU.

Microwave signals were sent from Earth stations to the probes,
which transmitted phase-locked signals back. Each station's signal
was derived from a hydrogen maser frequency reference with an
accuracy exceeding 1 part in $10^{12}$, and the Doppler-shifted
frequency of the returning signal was compared continuously.
Unlike the subsequent Voyager missions to the outer planets, the
Pioneers used spin stabilization, which maintains a spacecraft's
orientation without frequent use of thrusters. Consequently,
Doppler data was accumulated over long periods during which the
motions of the craft were undisturbed.

The Doppler data was checked against models of the probes' motions
by separate groups at NASA's Jet Propulsion Laboratory and The
Aerospace Corporation. \linebreak An unmodeled blueshift was found
in each case, equivalent to an acceleration $a_p$ of $\sim$$8\!
\times\! 10^{-8} \,{\rm cm/s}^2$. Extensive further analysis of
the Pioneer 10 data by Anderson, Laing, Lau, Liu, Neito and
Turyshev~\cite{alllnt} arrived at $(8.74 \pm 1.33) \!\times\!
10^{-8} {\rm cm/s}^2$, in the approximate direction of the Sun and
Earth. This effect hasn't been reconciled with general relativity.

According to general relativity, space-time is curved by
mass-energy.
 Quantum mechanics says space is filled with vacuum energy. Yet
measurements of the universe's large-scale curvature show none.
Wilczek~\cite{fw1} writes:
\begin{quote}
Any theory of gravity that fails to explain why our richly
structured vacuum, full of symmetry-breaking condensates and
virtual particles, does not weigh much more than it does is a
profoundly incomplete theory.
\end{quote}
And~\cite{fw2}
\begin{quote}
Since gravity is sensitive to all forms of energy it really ought
to see this stuff.\,.\,.\,. A straightforward estimation suggests
empty space should weigh several orders of magnitude (no misprint
here!) more than it does. It ``should" be much denser than a
neutron star, for example. The expected energy of empty space acts
like dark energy, with negative pressure, but there's much too
much of it. To me this discrepancy is the most mysterious fact in
all of physical science, the fact with the greatest potential to
rock the foundations.
\end{quote}

\nopagebreak A previous paper~\cite{kk} introduced an alternative to
general relativity, in which the gravitational motion of particles
and bodies is based on the optics of \nopagebreak de Broglie waves.
This ``weight problem" doesn't arise there, and inflation, strange
dark matter, and strange dark energy aren't needed for a flat
universe. The theory derives from general principles, but different
ones. Instead of introducing a geodesic principle, there is
Huygens', which is already part of quantum mechanics.

Poincar\'{e}'s principle of relativity is also extended to reference
frames in uniform gravitational potentials; the observed laws of
physics remain unchanged. There is no assumption of the equivalence
principle, in any of its various forms. Nevertheless, that principle
is effectively obeyed for weak fields, as shown for the lunar
orbit~\cite{kk}. And while neither Mach's principle nor manifest
covariance is assumed, as in general relativity, special relativity
is obeyed in uniform potentials.

While this quantum-mechanical theory agrees with the standard
experimental tests of general relativity, it makes different
predictions for the second-order solar deflection of starlight,
and for the pending NASA satellite experiment Gravity Probe
B~\cite{kk}.  After summarizing the basic theory, here we'll
derive it's prediction for the Pioneer probes.

General relativity is often characterized as a ``complete" theory
of gravity, but provides no way to distinguish the future from the
past. This acknowledged ``problem of time" is targeted in theories
of loop quantum gravity. Their intent is to make time as we
experience it completely unnecessary -- to make it go away.

Prigogine~\cite{ip} argued we can't, that time with a direction is
essential for describing thermodynamics and quantum mechanics. For
that purpose, theories of parameterized quantum mechanics introduce
a time parameter $\tau$ into Minkowski space-time, as a function of
a Newtonian time $t$. However Hartle~\cite{jbh} has noted it may be
impossible to do that to the curved space-time of general
relativity.

This gravity theory is built on the preferred-frame special
relativity advocated by Lorentz, Poincar\'{e}, and more recently by
Bell~\cite{jsb}. There space and time are kept separate, time is
already a Newtonian parameter, and nothing more is needed to
describe time's direction. Bell saw this relativity as a likely
necessity for a causal quantum mechanics. And it can be argued a
preferred frame is precisely identified by the universal cosmic
microwave background.

Instead of curving space-time, the fundamental effect of
gravitational potentials in this theory is a slowing of
quantum-mechanical waves. Where Einstein assumed an absolute speed
of light, with space and time variable, the assumption here is the
opposite. For an absolute space and time, of course the natural
coordinates are isotropic, and are the type we'll be using.

Gravitational potentials are treated as attributes of elementary
particles, having the same relativistic form as the
electromagnetic scalar potential.  In rectangular coordinates, the
potential due to a particle at the origin moving in $x$ is
\begin{equation}
\Phi \:=\, \frac{-\;G\,m_0}{\sqrt{\,x^2 + (y^2
+z^2)(1-v^2/\,c^2)}}
\end{equation}
where the role analogous to charge is played by its rest mass
$m_0$, and $G$ is the gravitational constant. (One gravitational
potential, vs. {\em ten} in general relativity.) There is also a
relativistic wave equation corresponding to that for the
electromagnetic scalar potential,

\begin{equation}
\nabla^2 \Phi \,-\, \frac{1}{c^2}\frac{\partial^2\Phi}{\partial
t^2} \:=\: 4 \pi G \rho
\end{equation}
where $\rho$ is the rest mass density.

Following are five basic transformations, whose
derivations from general principles can be found in the previous
paper~\cite{kk}. The speed of light as a function of the
gravitational potential varies as
\begin{equation}
c \:=\: c_0 \, e^{2 \Phi / c_0^2} \label{eq:3}
\end{equation}
where $c_0$ is the value in the absence of a gravitational
potential.  The velocity of de Broglie waves is similarly
\begin{equation}
V \:=\: V_0 \, e^{2 \Phi / c_0^2}
\end{equation}
where the $0$ subscript again indicates the corresponding quantity
with no potential.

The de Broglie frequency is
\begin{equation}
\nu \:=\: \nu_0 \, e^{\Phi /c_0^2} \label{eq:5}
\end{equation}
and the wavelength is
\begin{equation}
\lambda \:=\: \lambda_0 \, e^{\Phi/c_0^2} \label{eq:6}
\end{equation}
Since the rate of any clock is determined by the de Broglie
frequency of its particles, clocks in gravitational potentials
slow by the factor $e^{\Phi/c_0^2}$. The dimensions of atoms and
meter sticks are determined by the de Broglie wavelength, and
shrink by the same factor.

A particle or body's rest mass $m_0$ varies as
\begin{equation}
m_0 \:=\: m_{00} e^{-3 \Phi/c_0^2} \label{eq:57}
\end{equation}
where $m_{00}$ is the mass for a zero velocity and zero potential.
Another equation derived gives the relativistic acceleration
$\bf{a}$ of a particle or body
\begin{equation}
{\bf a} \:=\: -\nabla\Phi \left(e^{4\Phi/c_0^2}+ \frac
{\,v^2}{\,c_0^2}\right) + \frac{\,4{\bf v}}{\,c_0^2} \left(\frac
{\,d{\Phi}}{\,dt}\right) \label{eq:8}
\end{equation}
where $\bf{v}$ is its velocity.

Since there is no instability in the universe's geometry, there is
no Big Bang. But inherent instability exists in the gravitational
potential of the universe, and the overall speed of
light~\cite{kk}. The rate of change for the cosmological potential
is taken to be
\begin{equation}
\frac{\,d}{\,dt}\! \left( \frac{\,\Phi}{\,c_0^2} \right) \,=\: -H
\label{eq:9}
\end{equation}
where $H$ is the Hubble constant.  From this and Eq.~(\ref{eq:5}),
clocks are slowing. And from Eq.~(\ref{eq:6}), atoms and meter
sticks are shrinking.

The term ``relativity" was Poincar\'{e}'s, and to illustrate the
principle~\cite{hp} he asked: What if you went to bed one night,
and when you awoke the next day everything in the world was a
thousand times bigger? Would you notice anything?  As he pointed
out, such effects aren't observed locally, since measuring devices
change with the objects they measure. (According to
Einstein~\cite{ae},  he also insisted the true geometry of the
universe is Euclidean.)

Still, in this gradually evolving universe, an apparent expansion
would be seen in distant galaxies. Suppose a galaxy's light takes
time $t$ to reach an observer here. During that time, the
cosmological potential changes by $-H t $, and from
Eq.~(\ref{eq:3}), the speed of light diminishes by $e^{-2 H t}$.
Since no difference arises in the relative velocities of two
successive wavefronts, the absolute wavelength of light from a
remote source doesn't change after its emission.

But the wavelength of a local spectral reference shrinks, in
accord with Eq.~(\ref{eq:6}). We'll define the wavelength of an
observer's spectral reference as $\lambda_0$.  The quantity in
Eq.~(\ref{eq:6}) corresponding to $\Phi/c_0^2$ then is $+H t$, and
the relative wavelength $\lambda$ seen by the observer is
\begin{equation}
\lambda \:=\: \lambda_0 e^{H t} \:\cong\: \lambda_0 \left ( 1 + H
t \right )
\end{equation}
When the galaxy's distance $d$ is modest, this gives the familiar
Hubble relation for low redshifts,
\begin{equation}
\lambda \:\cong\: {\lambda_0}\,(\,1+H d/c\,)
\end{equation}
usually interpreted as an expansion of the universe.

The diminishing speed of light does change the absolute frequency
at which wavefronts arrive, and the apparent frequency of the
source. With $c$ slowing as $e^{-2 H t}$, the observed frequency
$\nu$ is
\begin{equation}
\nu \:=\: \nu_0 e^{- H t} \:\cong\: \nu_0 \left ( 1 - H t \right )
\end{equation}
where $\nu_0$ is the frequency of a local reference.  From
Eq.~(\ref{eq:5}), the frequency of light from a remote galaxy is
$e^{H t}$ greater than  $\nu_0$ when emitted.  But it's less by the
same factor when observed, remaining proportional to $c/\lambda$.

\section{Signal for a Stationary Earth and Non-gravitating Probe}

To calculate the predicted Pioneer Doppler signal, we'll start
with a simplified example, where there are no local gravitational
fields. Earth will be treated as stationary and massless, with the
probe moving away in a straight line. While the signal the probe
returns is is sent at a different frequency than the one it
receives, the two are phase-locked. So we'll treat these as a
simple reflection of the same signal by a moving body.

From the cosmological redshift described above, without a Doppler
shift, the relative frequency of the returning signal would be
\begin{equation}
\nu  \:\cong\: \nu_0 \left ( 1 - 2 H t \right )
\end{equation}
where $t$ is the signal's one-way travel time. After including the
first-order Doppler effect, this becomes
\begin{equation}
\nu  \:\cong\: \nu_0 \left (1 - \frac{\,2 \,v}{\,c} - 2 H t
\,\right ) \label{eq:14}
\end{equation}
where $v$ is the probe's velocity.

From Eq.~(\ref{eq:9}) and the final term in Eq.~(\ref{eq:8}), the
probe has a velocity-dependent acceleration
\begin{equation}
a \:=\: -4 H v  \label{eq:15}
\end{equation}
Where $T$ is the time since its departure, $v_i$ is its initial
velocity, and $\bar{v}$ the average, the probe's velocity
diminishes as
\begin{eqnarray}
v & \!= & \!v_i - 4\, \bar{v} H T \nonumber\\ & \!\cong &  \!v_i
(1 - 4 H T) \label{eq:16}
\end{eqnarray}

From Eq.~(\ref{eq:3}), the speed of light also changes as
\begin{eqnarray}
c & \!= & \!c_i e^{-2 H T} \nonumber\\ & \!\cong &  \!c_i (1 -\, 2
H T )  \label{eq:17}
\end{eqnarray}
with $c_i$ the initial speed of light when the probe departs
Earth. Since the terms involving $ H T$ are small, from the last
two equations, the ratio $v/c$ can be approximated as
\begin{equation}
\frac{\,v}{\,c}  \:\cong\, \frac{\,v_i}{\,c_i}\left (1 -\, 2 H T
\right)
\end{equation}

The probe's travel time and that for the signal are related
approximately by
\begin{equation}
T  \:\cong\, \frac{\,c_i \,t}{\,v_i}
\end{equation}
After substituting for $T$, the previous equation becomes
\begin{equation}
\frac{\,v}{\,c}  \:\cong\, \frac{\,v_i}{\,c_i} -\, 2 H t
\end{equation}
Then substituting for $v/c$ in Eq.~(\ref{eq:14}) gives
\begin{equation}
\nu  \:\cong\: \nu_0 \left (1 - \frac{\, 2 v_i}{\,c_i} + 2 H t
\,\right ) \label{eq:21}
\end{equation}

From the usual model of the probe's motion, where $v$ and $c$
remain equal to $v_i$ and $c_i$ respectively, the result is an
unmodeled blueshift of $2\nu_o H t$. (Note this approximation
doesn't hold for large values of $H t$, where $v$ and $c$ are
changing substantially.)

Anderson {\em et al.}~\cite{alllnt} discuss the possibility the
``Pioneer effect" is due to an extra acceleration $a_p$ of the
probes toward the Sun or Earth. Equating the blueshift from such
an acceleration to that from the last equation
\begin{equation}
\nu_0 \frac{\,2 a_p t}{c} \:=\: \nu_0 (2 H t)
\end{equation}
where the direction of $a_p$ is defined oppositely to the
acceleration $a$ above. Solving for $a_p$

\begin{equation} a_p \:=\: H c
\end{equation}
Anderson {\em et al.} note that various workers have observed the
value of $a_p$ is close to $c$ multiplied by the estimated Hubble
constant. The predicted blueshift here is equivalent to such an
acceleration.

\section{Earth's Orbit}

At around 29.8 km/s, Earth's orbital velocity with respect to the
solar system barycenter is more than twice that of the Pioneer 10
and 11 probes (now moving at 12.2 and 11.6 km/s respectively). Are
the Pioneer Doppler signals also influenced by change in Earth's
motion? To answer that, we'll need a description of its orbit in
an evolving universe.

If the Sun-Earth distance is regarded as a measuring rod, and the
frequency at which Earth orbits as a clock, from Eqs.~(\ref{eq:6})
and~(\ref{eq:5}), these quantities should diminish
relativistically as $e^{-H t}$.  Earth's average orbital speed is
proportional to the product, and should slow as $e^{-2 H t}$, by
the same factor as light.  As seen in Eqs.~(\ref{eq:16})
and~(\ref{eq:17}), the velocity of the probe in the preceding
example diminishes by about twice that factor.

Unlike that body, Earth is bound in a nearly circular orbit. Using
the same acceleration equation that was applied to the probe, we'll
show Earth's motion is approximately relativistic, slowing in
proportion to light. To do that, we'll first determine what Earth's
acceleration would be if it moves relativistically, and then compare
that to the acceleration given by Eq.~(\ref{eq:8}).

What we need are only small corrections to Earth's Newtonian
motion. And since the orbital radius varies by only $\pm 1.7$
percent, here we'll approximate the observed orbit as circular and
centered on the Sun. In absolute coordinates, the radius required
for a relativistic orbit then varies as
\begin{equation}
r  \:=\: r_0 e^{-H t} \label{eq:24}
\end{equation}
Also, the corresponding orbital velocity changes as
\begin{equation}
v  \:=\: v_0 e^{-2H t} \label{eq:25}
\end{equation}

Using $x$-$y$ coordinates, we'll place the Sun at the origin and
Earth at $(0,r)$, moving in approximately the +x direction. Taking
the derivative of Eq.~(\ref{eq:24}), \linebreak Earth's velocity
$v$ also has a component in the $y$ direction
\begin{eqnarray}
v_y & \!= & \!\frac{\,d r}{\,d t} \nonumber\\ & \!\cong &  -H r
\label{eq:26}
\end{eqnarray}
And the trajectory spirals inward at a very small angle $\alpha$
\begin{equation}
\alpha  \:\cong\: \frac{\,v_y}{\,v_x} \:\cong\: -\frac{H r}{v}
\end{equation}

With Earth's centripetal acceleration perpendicular to this angled
trajectory, it has a small component in the $x$ direction. If
Earth's speed were constant, its acceleration in that dimension
would be
\begin{equation}
a_x  \:\cong\: \frac{\:v^2}{\,r}\, \sin \alpha \:\cong\: - H v
\end{equation}
since $\alpha$ and sin $\alpha$ are effectively the same. That
deceleration corresponds to the orbit's increasing curvature. In
addition, Earth's orbital motion is slowing. The derivative of
Eq.~(\ref{eq:25}) gives
\begin{equation}
\frac{\,d v}{\,d t}  \:\cong\: - 2H v
\end{equation}
This deceleration is almost entirely in the $x$ dimension.
Combining this and the $x$ component of the centripetal
acceleration, the total $x$ deceleration is
\begin{equation}
a_x  \:\cong\: - 3 H v \label{eq:31}
\end{equation}

Now we'll compare this to the acceleration given by
Eq.~(\ref{eq:8}). From the Sun's potential gradient, the first term
of that equation gives an acceleration in the $-y$ direction. This
approximately equals Earth's centripetal acceleration.  For a spiral
trajectory with the same inward angle $\alpha$, it has a small
component in the direction of Earth's velocity vector.  That
acceleration $a_{\bf v}$ can be approximated in terms of the
centripetal acceleration as
\begin{equation}
a_{\bf v}  \:\cong\: -\frac{\:v^2}{\,r}\, \sin \alpha \:\cong\: H
v
\end{equation}
again substituting $\alpha$ for sin $\alpha$.

From the changing cosmological potential, the last term in
Eq.~(\ref{eq:8}) contributes a deceleration in that direction equal
to $-4 H v$. The combined effect is then
\begin{equation}
a_{\bf v}  \:\cong\: - 3 H v
\end{equation}
Since $a_{\bf v}$ is almost entirely in the $x$ direction, this also
gives Eq.~(\ref{eq:31}).

Again, that equation corresponds to an orbital velocity which
decreases as $e^{-2 H t}$, in proportion to the speed of light.
Earth's deceleration due to the changing cosmological potential is
partly manifested as change in its direction of motion. And its
deceleration is partially counteracted by the Sun's gravity, as the
planet spirals inward and gains kinetic energy.

Earth's orbit shrinks with Earth itself, as in a diminishing
Poincar\'{e} world.  And its orbital frequency slows together with
an atomic clock.  So there is no measured drift in the length of a
year.

\section{Predicted Signal}

To determine the contribution of Earth's motion to the Pioneer
signal, we'll suppose Earth orbits while the probe is stationary
at some large distance in the heliocentric reference frame.
Eq.~(\ref{eq:14}) applies again.  However, for an apparently
circular orbit, $v$ and $c$ diminish at the same rate and the
ratio $v/c$ doesn't change.  Hence the decrease in Earth's
tangential velocity doesn't contribute a Doppler shift.

The inward radial component of Earth's velocity described by
Eq.~(\ref{eq:26}) does give a Doppler shift.  This velocity varies
sinusoidally in the probe's direction, and the resulting shift can
be expressed as
\begin{equation}
\Delta \nu  \:\cong\: -2 \bar{\nu} \,\frac{ H\, r \cos \theta}{c}
\end{equation}
where  $\bar{\nu}$ is the average observed frequency, $r$ the
radius of Earth's orbit and $\theta$ is the angle between the
heliocentric position vectors for the probe and Earth.

The last term in Eq.~(\ref{eq:14}) also contributes.  Due to Earth's
changing position, the signal's one-way travel time $t$ varies by
$(-r \cos \theta / c)$.  The resulting shift is the opposite of that
in the last equation, with the redshift arising when Earth is
farther from the probe.  Consequently, Earth's motion brings no net
unmodeled frequency shift.  We'll express that as
\begin{equation}
\Delta \nu_e  \,\cong\: 0
\end{equation}

Now we'll determine the unmodeled shift due to the probe's motion
on its actual trajectory, again in the heliocentric frame.  First
we'll generalize Eq.~(\ref{eq:21}) to give the unmodeled shift for
an arbitrary, short segment of the probe's trajectory. We rewrite
that equation as
\begin{equation}
\nu  \:\cong\: \nu_i \left (1 - \frac{\, 2 v_i}{\,c_i} + 2 H t
\,\right ) \label{eq:34}
\end{equation}
where $\nu_i$ is the observed frequency when the probe is at the
beginning of a segment, $v_i$ and $c_i$ are the quantities at that
point, and $t$ is the additional time a signal takes to reach the
probe after it moves beyond that point. Since we're not
calculating the Doppler shift due to the probe's modeled
acceleration by local gravitational fields, those are neglected
again.

The derivation of this equation is the same given in Section 2 for
Eq.~(\ref{eq:21}).  In this case, $T$ refers to the probe's travel
time from the beginning of the trajectory segment. Since the
Doppler and cosmological shifts are functions of its radial
velocity, we define the quantities $v$ and $v_i$ as the radial
velocity components. Likewise, $a$ in Eq.~(\ref{eq:15}) refers to
its radial acceleration. After these definitions, the previous
derivation holds when the probe's net motion is in any direction.

A sufficiently short segment of the probe's curved trajectory can be
represented as a straight line, for which Eq.~(\ref{eq:34}) gives an
unmodeled shift of $2 \nu_i H t$.  Summing the shifts for successive
segments, in the limit where the segment length goes to zero, gives
the total.  The resulting unmodeled frequency shift for an arbitrary
trajectory can again be expressed as
\begin{equation}
\Delta \nu_p  \:\cong\: 2 \nu_0 H t
\end{equation}
where $\nu_0$ is the reference and $t$ is the total one-way signal
travel time.

Note this equation doesn't depend on the probe's velocity -- since
$t$ is proportional to its radial distance traveled, not the time it
takes to reach a given radius.  As Anderson {\em et. al.} point out,
rather than an acceleration of the probe, this effect could instead
be attributed to a slowing frequency reference on Earth. (In that
case it may be more apparent the resulting frequency shift is a
function of the probe's distance.)

As shown previously, the blueshift given by the last equation is
equivalent to the anomalous acceleration $a_p$ of Anderson {\em et.
al.}, when its value is $H c$.  Several effects are involved here: a
diminishing speed of light, a slowing frequency reference on Earth,
and a four-times-larger deceleration of the probe.

(Although $a_p$ was taken to be directed toward the Sun, the $-4 H
v$ acceleration here is opposite the probe's motion. Its main
component is toward the Sun, and contributes to a net frequency
shift equivalent to $a_p$.  It also has a small perpendicular
component. That causes no additional Doppler effect, but shifts the
probe's angular position in the heliocentric reference frame. The
contribution of Earth's orbital motion to the Pioneer signal then
changes, in both phase and amplitude. This may relate to a small
annual oscillation found by Anderson {\em et. al.}~\cite{alllnt}.)

Equating the value of $a_p$ to $H c$, $H$ is then $89.8 \pm 13.6$
km/s/Mpc.  Measurements of the Hubble constant from astronomical
observations have yielded widely varying results. With the goal of
measuring $H$ to ten percent accuracy, Freedman {\em et.
al.}~\cite{wlf} have used the Hubble Space Telescope to calibrate
the distances of Cepheid variable stars in nearby spiral galaxies.
Their final estimate was $72 \pm 8$ km/s/Mpc.

Using very long baseline interferometry, Herrnstein {\em et.
al.}~\cite{jrh} have made a precise, direct measurement of the
distance to a water maser in one of the same galaxies. It puts the
galaxy, NGC4258, twelve percent closer.  And when the Cepheid
yardstick is recalibrated accordingly, the result is a Hubble
constant of $80 \pm 9$ km/s/Mpc.  The value of $H$ given by this
theory and the Pioneer data is in agreement with both these
estimates.

\section{Conclusions}

In a {\em New York Times} essay commemorating a century of quantum
mechanics, John Wheeler \cite{jaw} remarks:

\begin{quote}
It was 228 years later when Einstein, in his theory of general
relativity, attributed gravity to the curvature of spacetime. . .
.  Even that may not be the final answer. After all, gravity and
quantum mechanics have yet to be joined harmoniously.
\end{quote}

General relativity fails to explain why the universe isn't curved
by the vacuum's quantum-mechanical energy.  Since this alternative
is based on de Broglie waves instead of space-time curvature,
there is no need to reconcile it with quantum mechanics or the
observed flatness of the universe. There also is no ``problem of
time." Further, unlike general relativity, this theory agrees with
the signals from Pioneer 10 and 11.


\section*{Appendix: Elliptical Earth Orbit}

Treating Earth's observed orbit as circular, we found its motion
contributes no \linebreak unmodeled frequency shift to the Pioneer
Doppler signal. Here we'll explore the effects of its orbital
eccentricity $\epsilon$, which is 0.0167.  (The terms ``Earth" and
``Sun" are used loosely to refer to the barycenters of the
Earth-Moon and Solar systems.)

We'll use $x$-$y$ coordinates, where the origin corresponds to both
the Sun's position and a focus of Earth's elliptical orbit.
Initially, the aphelion lies on the positive $x$ axis, and
perihelion on the negative. For an ellipse of small eccentricity,
the major and minor axes are approximately equal. Consequently,
where $\bar{r}$ is Earth's mean orbital radius, its trajectory can
be approximated as a circle with radius $\bar{r}$, whose center lies
on the $x$ axis, displaced a positive distance $\epsilon \bar{r}$
from the origin.

We'll also use polar coordinates with the same origin, where Earth's
radial position is approximately
\begin{equation}
r \:\cong\: \bar{r} + \epsilon \bar{r} \cos \theta \label{eq:37}
\end{equation}
Its velocity in terms of the mean is given by
\begin{equation}
v \:\cong\: \bar{v} - \epsilon \bar{v} \cos \theta \label{eq:38}
\end{equation}
Substituting from
\begin{equation}
\theta \:\cong\: \frac{\,\bar{v} t}{\,\bar{r}}
\end{equation}
and taking the time derivative of the velocity gives a tangential
acceleration along the trajectory
\begin{equation}
a_t \:\cong\: \frac{\,\epsilon \bar{\,v}^2}{\,\bar{r}} \sin
\theta\label{eq:40}
\end{equation}

Our previous analysis omitted the varying velocity component
$(-\epsilon \bar{v} \cos \theta)$ in Eq.~(\ref{eq:38}). From the
changing cosmological potential given by Eq.~(\ref{eq:9}) and the
final term in Eq.~(\ref{eq:8}), this velocity component produces an
added tangential acceleration
\begin{equation}
a_t \:\cong\: 4 H \epsilon \bar{v} \cos \theta \label{eq:41}
\end{equation}

We'll find this alternating acceleration is offset by a $y$
displacement of Earth's approximately circular trajectory.  The
circle's center has the $y$ coordinate,
\begin{equation}
y_c \:\cong\: \frac {\,4 H \epsilon \bar{\,r}^{\,2}}{\,\bar{v}}
\end{equation}
Taking $H$ equal to 89.8 km/s/Mpc, this gives a $y$ shift of $14.6$
cm. Eq.~(\ref{eq:37}) for the orbital radius becomes
\begin{equation}
r \:\cong\: \bar{r} + \epsilon \bar{r} \cos \theta + \frac {\,4 H
\epsilon \bar{\,r}^{\,2}}{\,\bar{v}} \sin \theta
\end{equation}
By itself, the final term of this would introduce another velocity
component in Eq.~(\ref{eq:38}), with the value $(-4 H \epsilon
\bar{r} \sin \theta)$, and a tangential acceleration opposite that
in Eq.~(\ref{eq:41}). From the cancelation of those accelerations,
Earth's net velocity and acceleration are unchanged, and still
described by Eqs.~(\ref{eq:38}) and~(\ref{eq:40}).

Due to the shift of Earth's trajectory, its aphelion and perihelion
no longer coincide with the points of minimum and maximum orbital
velocity.  Since the trajectory is effectively circular, the
aphelion and perihelion lie on a line passing through its center and
the Sun's position.  The line's angle $\beta$ with respect to the
$x$ axis is given approximately by its slope
\begin{equation}
\beta \:\cong\: \frac{\,y_c}{\,\epsilon \bar{r}} \:\cong\: \frac
{\,4 H \bar{r}}{\,\bar{v}}
\end{equation}
where $\beta$ is in radians.

This is the angular difference between the aphelion and the point of
minimum velocity in the heliocentric reference frame. Over the
entire orbit, the magnitude of Earth's velocity is shifted by this
angle from its usual value. Using the same $H$, \linebreak $\beta$
is $5.8 \times 10^{-11}$ radians, or $12$ $\mu$arcsec. Multiplying
$\beta$ by $\bar{r}$, this corresponds to a distance of 8.7 meters
along Earth's trajectory. Dividing that by $\bar{v}$, the magnitude
of its velocity precedes the usual in phase by about $0.29$
millisec.

This perturbation of Earth's orbit appears too small to be detected
in the Pioneer Doppler signals.  However, its effects are much
larger for the outer planets, and may be observable in angular
position measurements.  For Pluto, $\beta$ is 3 milliarcsec.

\vfill\eject


\begin{thebibliography}{xx}
\bibitem{kk}
\newblock{K. Krogh, ``Gravitation without curved space-time," astro-ph/9910325.}
\bibitem{alllnt}
\newblock{J. D. Anderson, P. A. Laing, E. L. Lau, A. S. Liu, M. M. Neito and S. G. Turyshev,``Study of the anomalous acceleration of Pioneer 10 and 11," {\em Phys. Rev. D\,} \textbf{65}, 082004 (2002).}
\bibitem{fw1}
\newblock{F. Wilczek, ``Scaling Mount Planck III: Is that all there is?" {\em Physics Today},  \textbf{55}(8), 10 (2002).}
\bibitem{fw2}
\newblock{F. Wilczek, ``The universe is a strange place," astro-ph/0401347.}
\bibitem{ip}
\newblock{I. Prigogine, {\em The End of Certainty}, (The Free Press, 1997).}
\bibitem{jbh}
\newblock{J. B. Hartle, ``Time and time functions in parameterized non-relativistic quantum mechanics," {\em Class. Quantum Grav.} \textbf{13}, 361 (1996).}
\bibitem{jsb}
\newblock{J. S. Bell, ``How to teach special relativity," {\em Speakable and Unspeakable in Quantum Mechanics}, pp. 67-80 (Cambridge University Press, 1987).}
\bibitem{hp}
\newblock{H. Poincar\'{e}, {\em Science et M\'{e}thode}, (E. Flammarion, Paris, 1908).}
\bibitem{ae}
\newblock{A. Einstein, ``Geometry and Experience," {\em Sidelights on Relativity}, pp. 33-39 (Dover, 1922).}
\bibitem{wlf}
\newblock{W. L. Freedman {\em et al.}, ``Final results from the Hubble Space Telescope Key Project to measure the Hubble constant," {\em Astrophys.J.} \textbf{553}, 47 (2001).}
\bibitem{jrh}
\newblock{J. R. Herrnstein {\em et al.}, ``A geometric distance to the galaxy NGC4258 from orbital motions in a nuclear gas disk," {\em Nature} \textbf{400}, 539 (1999).}
\bibitem{jaw}
\newblock{J. A. Wheeler, ``A practical tool that is puzzling as well," {\em New York Times} December 12, D1 (2000).}

\end{thebibliography}
\end{document}